\documentstyle[aps,prl]{revtex}
\begin{document}
\twocolumn[\hsize\textwidth\columnwidth\hsize\csname
@twocolumnfalse\endcsname

\title{Extremal Optimization of Graph Partitioning at the Percolation Threshold}

\author{Stefan Boettcher$^{1,2}$}
\address{
$^1$Physics Department, Emory University, Atlanta, Georgia 30322, USA\\
$^2$Center for Nonlinear Studies, Los Alamos National Laboratory, Los
Alamos, NM 87545, USA}

\date{\today}
\maketitle

\begin{abstract}
The benefits of a recently proposed method to approximate hard
optimization  problems are demonstrated on the graph partitioning
problem. The performance of this new method, called Extremal
Optimization, is compared to  Simulated Annealing in extensive
numerical simulations.  While generally a complex (NP-hard) problem,
the optimization of  the graph partitions is particularly difficult
for sparse graphs with average connectivities near the percolation
threshold.  At this threshold, the relative error of Simulated
Annealing for large graphs is found to diverge relative to
Extremal Optimization at equalized runtime. On the other hand, Extremal
Optimization,  based on the extremal dynamics of self-organized
critical systems, reproduces known results about optimal partitions at
this critical point quite well.
\medskip
\end{abstract}
]

\section{Introduction}
\label{intro}
The optimization of systems with many degrees of freedom with respect
to some cost function is a frequently encountered task in physics and
beyond \cite{MPV}.  In cases where the relation between individual
components of the system is frustrated \cite{Toulouse}, such a cost
function often exhibits a complex ``landscape'' \cite{cnls} over the
space of all configurations. For growing system size, the cost function
may exhibit an exponentially increasing number of unrelated local
extrema separated by sizable barriers which makes the search for the
exact, optimal solution usually unreasonably costly.  Thus, it is of
great importance to develop fast and reliable methods to find
near-optimal solutions for such problems.

The observation of certain physical processes, in particular the
annealing of disordered materials, have lead to general-purpose
optimization methods such as ``simulated annealing'' (SA) \cite{Science,Cerny}.
SA applies the formalism  of equilibrium statistical mechanics and in principle only requires the cost function as input. Thus, it is applicable to a variety of problems.
But the performance of SA is hard to assess in general, even when
limited to the standard combinatorial optimization problems. Aside
from a multitude of adjustable parameters that crucially determine the
quality of SA's performance in a particular context, typical
combinatorial optimization problems themselves possess various
parameters that may change the landscape and SA's behavior
drastically \cite{Sorkin}.

In this paper we will explore the properties of a new general-purpose
method, called Extremal Optimization (EO) \cite{BoPe1}, in
comparison with SA. In contrast to SA, EO is based on ideas from non-equilibrium physics. As the basis for comparison we will use the graph partitioning problem (GPP), a standard NP-hard combinatorial
optimization problem \cite{G+J} with similarities to disordered spin systems. We find that the GPP has a critical
point as a function of the connectivity of graphs, with a less complex
phase at lower connectivities. This critical point is related to the
percolation transition of the graphs. Near this critical point, the
performance of SA markedly deteriorates while EO produces only small
errors.

This paper is organized as follows: In the next section we describe the 
philosophy behind the EO method, in Sec. \ref{GPP} we introduce the graph
partitioning problem, and Sec. \ref{NumEx} we present the algorithms and the 
results obtained in our numerical comparison of SA and EO, followed by 
conclusions in Sec. \ref{conclusions}.

\section{Extremal Optimization}
\label{EOintro}
EO provides an entirely new approach to optimization \cite{BoPe1},
based on the non-equilibrium dynamics of systems exhibiting
self-organized criticality (SOC)\cite{BTW}. SOC often emerges when a
system is dominated by the evolution of extremely atypical degrees of
freedom \cite{PMB}. 

A simple example of such a dynamical system which inspired the development of EO is the Bak-Sneppen model \cite{BS}. There,
species 
are represented by a number between 0 and 1 that indicates their
``fitness,'' located on the sites of a lattice. The smallest number
(representing the worst adapted species) at each update is discarded
and replaced with a new number drawn from a uniform distribution on
$[0,1]$.  Without any interactions, all the numbers in the system
would eventually become 1.  But obvious interdependencies between species
provide constraints for balancing the systems' fitness with that of
each species: The change of fitness in one species impacts the fitness
of an interrelated species. In the Bak-Sneppen model, the fitness
values on all sites neighboring the smallest number at that time step
are simply replaced with new random numbers as well \cite{REM1}. After
a certain number of such updates, the system organizes itself into a
highly correlated state known as self-organized criticality (SOC)
\cite{BTW}. 

In the SOC state, almost all species have reached a fitness
above a certain threshold. But these species merely possess what is referred to as
punctuated equilibrium \cite{BS,G+E}, because the co-evolutionary activity
is bound to return in a chain reaction where a weakened neighbor can
undermine one's own fitness. Fluctuations that rearrange the fitness of many
species abound and can rise to the size of the system itself,
making any possible configuration accessible. Hence, such non-equilibrium systems provide a high degree of adaptation 
for most entities in the system without limiting the scale of change towards even better states. 

\begin{figure}
\vskip 4.40truein \includegraphics{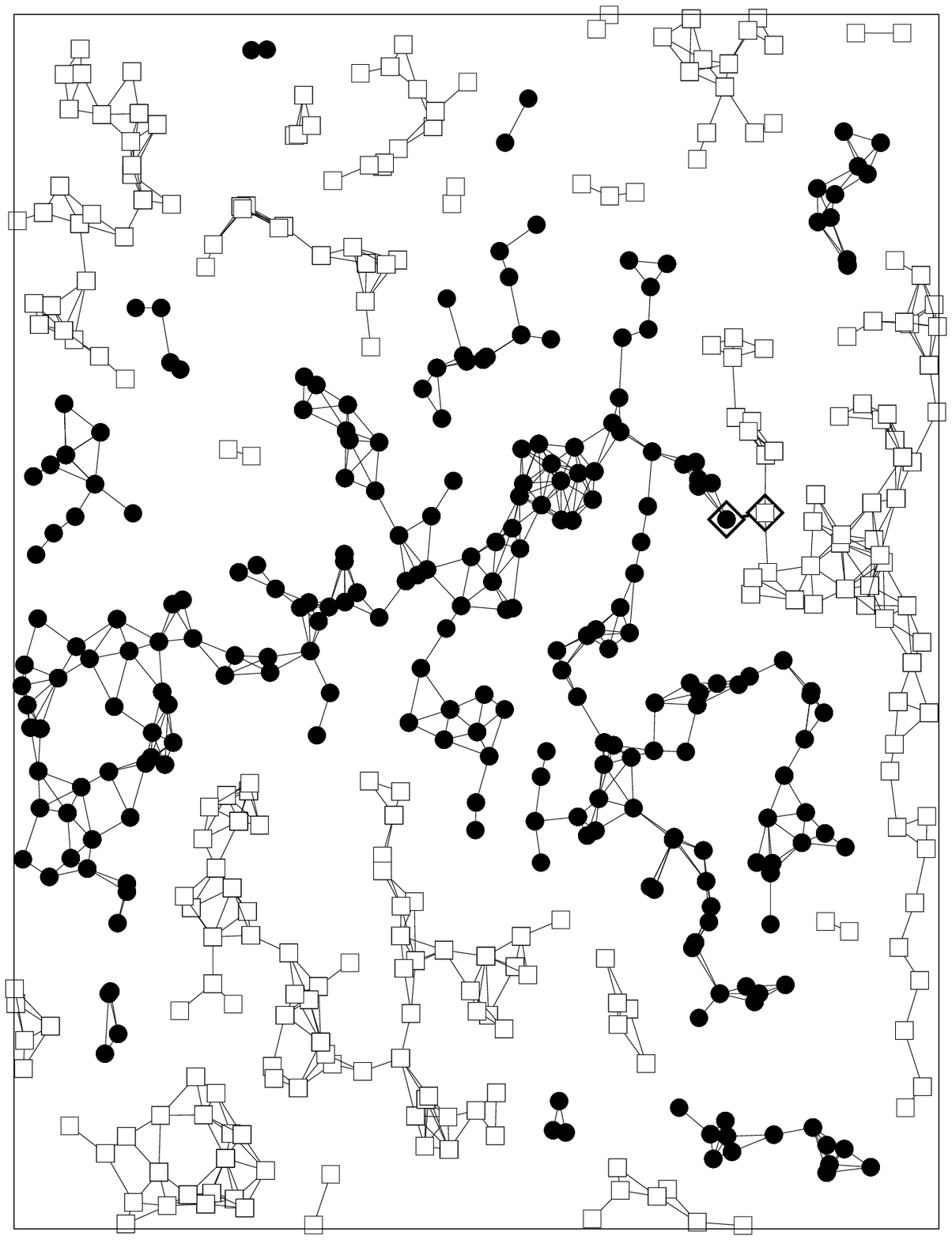} \includegraphics{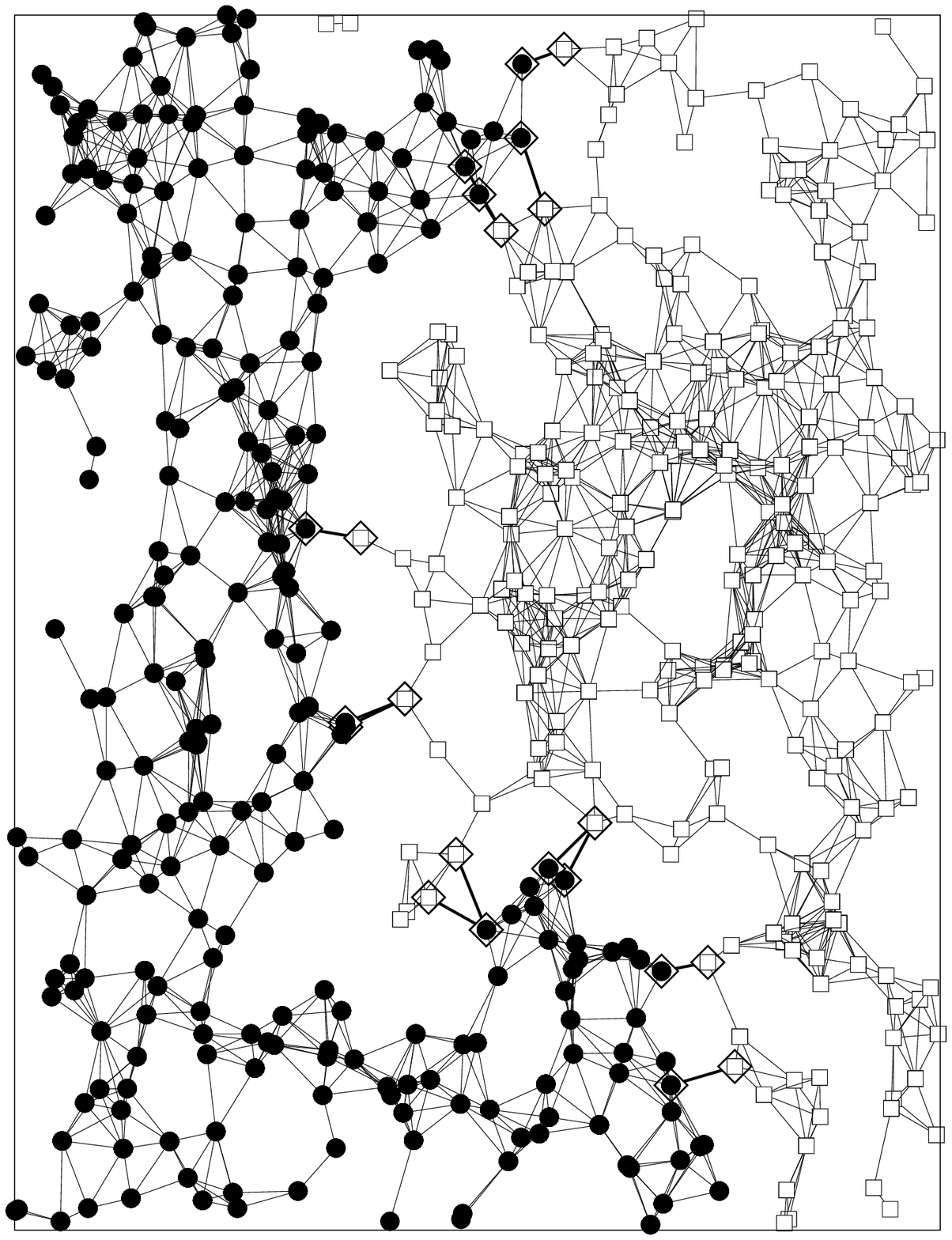}
\caption{Two random geometric graphs, $N=500$, with $\alpha=4$ (top) and
$\alpha=8$ (bottom) in an optimized configuration found by EO. At $\alpha=4$ the
graph barely percolates, with merely one ``bad'' edge (between points
of opposite sets, masked by diamonds) connecting a set of 250 round
points with a set of 250 square points, thus $m_{\rm opt}=1$. For the
denser graph on the bottom, EO reduced the cutsize to $m_{\rm opt}=13$.
}
\label{geograph}
\end{figure} 

EO attempts to utilize this phenomenology to obtain near-optimal solutions for optimization problems \cite{GECCO}. For instance, in a spin glass system \cite{MPV} we may consider as
fitness for each spin its contribution to the total energy of the
system. EO would search for ground state configurations by perturbing
preferentially spins  with large contributions. Like in the
Bak-Sneppen model, such perturbations  would be local, random rearrangements of those poorly adapted spins, allowing for
better as well as for worse outcomes at each update. In the same way
as systems exhibiting SOC get driven recurrently towards a small subset of attractor
states through a sequence of ``avalanches'' \cite{Dhar,BTW}, EO can
fluctuate widely to escape local optima while the extremal selection
process ensures recurrent approaches to many near-optimal
configurations. Especially in exploring low temperature properties of
disordered spin systems, those qualities may help to avoid the
extremely slow relaxation behavior faced by heat bath based
approaches \cite{Binder}. In that, EO provides an approach alternative -- and
apparently equally capable \cite{BoPe1} -- to Genetic Algorithms,
which are often the only means to illuminate those important properties
\cite{HO}. The partitioning of sparse graphs as discussed here
is particularly pertinent in preparation for similar studies on actual
spin glasses.

It has been observed that many optimization problems exhibit critical
points that separate off phases with simple cases of a generally hard
problem \cite{cheese}. Near such a critical point, finding solutions
becomes particularly difficult for local search methods which proceed
by exploring for an existing solution some neighborhood in
configuration space. There, near-optimal solutions become widely
separated with diverging barrier heights between them. It is not
surprising that search methods based on heat-bath techniques like SA
are not particularly successful in this highly correlated state
\cite{Binder}. In contrast, the driven dynamics of EO does not possess
any temperature control parameters to successively limit the scale of
its fluctuations. Our numerical results in Sec. \ref{NumEx} show that EO's performance does not diminish near such a critical point. A non-equilibrium approach like EO may thus provide
a general-purpose optimization method that is complementary to SA: While SA has the advantage far from this critical point, EO appears to work well ``where the {\sl really} hard problems are'' \cite{cheese}.

\section{Graph Partitioning}
\label{GPP}
To illustrate the properties of EO and its differences with SA, we focus in this paper on the well-studied graph partitioning problem (GPP). In particular, we will consider the GPP near a phase transition where the optimization problem becomes especially difficult and possesses many similarities with physical systems.

\subsection{Formulation of the Problem}
The graph (bi-)partitioning problem is easy to formulate: Take $N$
points where $N$ is an even number, let any pair of two points be
connected by an edge with a certain probability, divide the points
into two sets of equal size $N/2$ such that the number of edges
connecting both sets, the ``cutsize'' $m$, is minimal: $m=m_{\rm
opt}$. The global constraint of an equal division of the points
between the sets places this problem generally among the hardest
problems in combinatorial optimization, requiring a computational
effort that would grow faster than any power of $N$ to determine the
{\sl exact} solution with certainty \cite{G+J}.  The two physically 
motivated optimization methods, SA and EO, which we focus on here,
usually obtain {\sl approximate} solutions in polynomial time.

For random graphs, the GPP depends on the probability $p$ with which
any two points in the system are connected.  Thus, $p$ determines the
total number of edges in an instance, $L=pN(N-1)/2$ on average, and
its mean connectivity per point, $\alpha=p(N-1)$ on
average. Alternatively, we can formulate a ``geometric'' GPP by
specifying $N$ randomly distributed points in the $2$-dimensional unit
square which are connected with each other if they are located within
a distance $d$ of one another. Then, the average expected connectivity
$\alpha$ of such a graph is given by $\alpha=N\pi d^2$. This form of
the GPP has the advantage of a simple graphical representation, shown in
Fig.~\ref{geograph}. 

It is known that geometric graphs are harder to optimize than random 
graphs \cite{Johnson}. The characteristics of the GPP for
random and geometric graphs at low connectivity appear to be very
different due to the dominance of long loops and short loops, resp.,
and we present results for both types of graphs here. In fact, in the
case of random graphs the structure is locally tree-like which allows
for a mean-field treatment that yields exact results \cite{ER,MP,JKLP}. In
turn, the geometric case corresponds to continuum percolation of
``soft'' (overlapping) circles for which precise numerical results
exist \cite{Balberg}. Finally, we also try to determine the average
ground state energy of a dilute ferro-magnetic system on a cubic
lattice at fixed (zero) magnetization, which amounts to the equal
partitioning of ``up'' and ``down'' spins while minimizing the
interface between both types \cite{MP}. Here, each vertex of the
lattice holds a $\pm$-spin, and any two nearest-neighbor spins either
possess a ferromagnetic coupling of unit strength or are unconnected. The probability that a coupling exists is fixed
such that the average connectivity of the system is $\alpha$.

\subsection{Graph Partitioning and Percolation}
Like many other optimization problems, the GPP exhibits a critical point 
as a function of its parameters \cite{cheese}. In case of the GPP we observe this critical point as a function of the connectivity $\alpha$ of graphs, with the cutsize $m_{\rm opt}$ as the order parameter. In fact, the critical point of partitioning is closely linked to the percolation threshold of graphs. In our numerical simulations we proceed by averaging over many instances of a class of graphs and try to reproduce well-known results from the corresponding percolation problem. Of course, using stochastic optimization methods (instead of
cluster enumeration) is neither an efficient nor a precise means to
determine percolation thresholds. But in turn we obtain also some
valuable information about the scaling behavior of the average cost $<m_{\rm opt}>$ for optimal
partitions near the threshold that goes beyond the
percolating properties of these graphs.

We note, in accordance with Ref.~\cite{F+A}, that the critical point separates  between hard cases and easy-to-solve cases of the GPP. The transition is related to the corresponding
percolation problem for the graphs in the following manner: If the
mean connectivity $\alpha$ is very small, the graph of $N$ points
consists mainly of disconnected, small clusters or isolated points
which can be enumerated and sorted into two equal partitions in
polynomial time with no edges between them ($m_{\rm opt}=0$). If
$\alpha$ is large and the probability that any two points are
connected is $p={\rm O}(1)$, almost all points are connected into one
giant cluster with $m_{\rm opt}={\rm O}(N^2)$, and almost any
partition leads to an acceptable solution. But when $p={\rm O}(1/N)$,
i.~e. $\alpha={\rm O}(1)$, the distribution of cluster sizes is broad,
and the partitioning problem becomes nontrivial.  Obviously, as soon
as a cluster of size $>N/2$ appears, $m_{\rm opt}$ must be
positive. In this sense, we observe for $N\to\infty$ a sharp, percolation-like transition at an
$\alpha_{\rm crit}$ with the cutsize $m_{\rm opt}$ as the order
parameter.

For random graphs it is known that a cluster of
size $N$ exists for $\alpha>1$ \cite{ER}, but only for
$\alpha>\alpha_{\rm c}=2\ln2\approx1.386$ do we find a cluster of size
$>N/2$ \cite{MP}. Geometric graphs in $D=2$ are known to percolate at
about $\alpha=4.5$ \cite{Balberg}, and we would expect $\alpha_{\rm
c}$ for the GPP to be slightly larger than that. Also, the dilute
ferro-magnet should exhibit a non-trivial energy when the fraction of
occupied bonds reaches slightly beyond the critical point $p_{\rm
c}\approx0.2488$ for bond percolation on a cubic ($D=3$) lattice \cite{S+A},
i. e. for connectivities $\alpha>2Dp_{\rm c}\approx1.493$.

\section{Numerical Experiments}
\label{NumEx}
\subsection{Simulated Annealing Algorithm}
\label{SAalgo}

In SA \cite{Science}, we try to minimize a global cost function given
by $f=m+\mu(P_1-P_2)^2$, where $P_1$ and $P_2$ are the number of
points in the respective sets.  Allowing the size of the sets to
fluctuate is required to improve SA's performance in outcome and
computational time at the cost of an arbitrary parameter $\mu$ to
be determined. Then, starting at a ``temperature'' $T_0$, the
annealing schedule proceeds with $lN$ trial Monte-Carlo steps on $f$
by tentatively moving a randomly chosen point from one set to the
other (which changes $m$) to equilibrate the system.  This move is
accepted, if $f$ improves or if the Boltzmann factor $\exp[(f_{\rm
old}-f_{\rm new})/T]$ is larger than a randomly drawn number between
$0$ and $1$. Otherwise the move is rejected and the process continues
with another randomly chosen point. After that, we set
$T_i=T_{i-1}(1-\epsilon)$, equilibrate again for $lN$ trials, and so
on, until the MC acceptance rate drops below $A_{\rm stop}$ for $K$
consecutive temperature levels. At this point the optimization process
can be considered ``frozen'' and the configuration should be
near-optimal, $m\approx m_{\rm opt}$ (and balanced, $P1=P2$). While SA is intuitive,
controlled, and of very general applicability, its performance in
practice is strongly dependent on the multitude of parameters which have
to be arduously tuned. For us it is thus expedient (and most
unbiased!) to rely on an extensive study of SA for graph
partitioning \cite{Johnson} which determined $\mu=0.05$, $T_0=2.5$,
$\epsilon=0.04$, $A_{\rm stop}=2\%$, and $K=5$. Ref.~\cite{Johnson}
set $l=16$, but performance improved noticeably for our choice,
$l=64$. 

\subsection{Extremal Optimization Algorithm}
\label{EOalgo}
In EO \cite{BoPe1}, each point $i$ obtains a ``fitness''
$\lambda_i=g_i/(g_i+b_i)$ where $g_i$ and $b_i$ are the number of
``good'' and ``bad'' edges that connect that point within its set and
across the partition, resp. (We fix $\lambda_i=1$ for isolated
points.) Of course, point $i$ has an individual connectivity of
$\alpha_i=g_i+b_i$ while the overall mean connectivity of a graph is
given by $\alpha=\sum_i\alpha_i/N$. The current cutsize is given by
$m=\sum_ib_i/2$. At all times, an ordered list
$\lambda_1\leq\lambda_2\leq\ldots\leq\lambda_N$ is maintained where
$\lambda_n$ is the fitness of the point with the $n$-th rank in the
list.

At each update we draw two numbers, $1\leq n_1,n_2\leq N$, from a
probability distribution 
\begin{eqnarray}
P(n)\sim n^{-\tau}.
\label{pdf}
\end{eqnarray}
Then we pick the points
which are elements $n_1$ and $n_2$ of the rank-ordered list of
fitnesses. (We repeat a drawing of $n_2$ until we obtain a point that
is from the opposite set than $n_1$.)  These two points swap sets {\it
no matter what} the resulting new cutsize $m$ may be, in notable
distinction to the (temperature-) scale-dependent Monte Carlo update
in SA. Then, these two points, and all points they are connected to
($2\alpha$ on average), reevaluate their fitness $\lambda$. Finally,
the ranked list of $\lambda$'s is reordered using a ``heap'' at a
computational cost $\propto \alpha\ln{N}$, and the process is started
again. We repeat this process for a number of update steps per run
that rises linearly with system size, and we store the best result
generated along the way. Note that no scales are
introduced into the process, since the selection follows a scale-free
power-law distribution $P(n)$ and since -- unlike in a heat bath --
all moves are accepted, allowing for fluctuations on all scales.
Instead of a global cost function, the rank-ordered list of fitnesses
provides the information about optimal configurations.  This
information emerges in a self-organized manner merely by selecting
with a bias {\it against} badly adapted points, instead of
``breeding'' better ones \cite{BS}.

There is merely one parameter, the exponent $\tau$ in the probability distribution in Eq.~(\ref{pdf}), that controls the selection process and optimizes the performance of EO. In initial studies, we determined $\tau=1.4$ as the optimal value for all graphs. It is intuitive that such an
optimal value of $\tau$ should exist: If $\tau$ is too small, points
would be picked purely at random with no gradient towards a good
partition, while if $\tau$ is too large, only a small number of points
with particularly bad fitness would be chosen over and over again,
confining the system to a poor local optimum. It is a surprising
numerical result that this value of $\tau$ appears to be rather
universal, independent of $N$, $\alpha$, and the type of graph considered.

\subsection{Testbed of Graphs}
\label{testbed}
In our numerical simulations we have generated random and $2D$
geometric graphs of varying connectivity by choosing $p$ or $d$,
resp. For any instance of a graph labeled by a ``connectivity
$\alpha$'', the actual connectivity not only varies from point to
point, but also the mean connectivity of such graphs follows a normal
distribution. (In particular for geometric graphs it is shifted to
lower values due to the loss of connectivity at the boundaries.)  For
$N=500$, 1000, 2000, 4000, 8000, and 16000, we varied the connectivity
between $\alpha=1.25$ and $\alpha=5$ for random graphs, and $\alpha=4$
and $\alpha=10$ for geometric graphs.  Then, for each $\alpha$ we
generated 16 different instances of graphs, identical for SA and
EO. On each instance, we performed 8 (32) optimization runs for random (geometric)
graphs, both for EO and SA. Each run, we used a
new random seed to establish an initial partition of the points. SA's
runs terminate when the system freezes.  We terminated EO-runs after
$200N$ updates, leading to a comparabile runtime between both methods.

\begin{figure}
\vskip 2.1truein 
\includegraphics{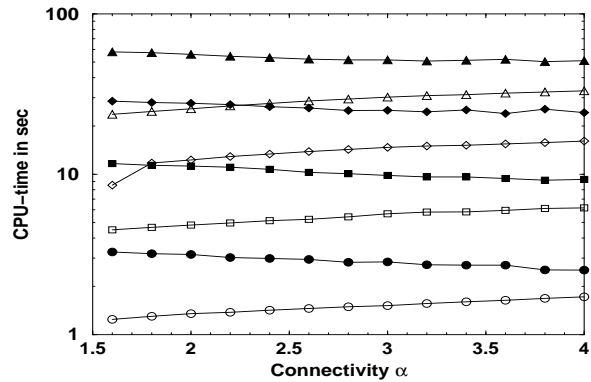}
\caption{Typical runtime comparison of SA and EO for the ferromagnet on a 
200MHz Pentium. Runtimes generally rise $\propto N$, fall with the 
connectivity $\alpha$ for SA (filled symbols), but rise $\propto \alpha$ for 
EO (unfilled symbols). Circles refer to an average runtime of graphs with 
$N=512=8^3$ points, squares to $N=1728=12^3$, diamonds to $N=4096=16^3$, and 
triangles to $N=8000=20^3$.
}
\label{runtime}
\end{figure}

For the dilute ferro-magnet, we fixed the number of couplings to
obtain a specific average connectivity $\alpha$. Those couplings were
then placed on random links between nearest-neighbor spins to generate an
instance. We used 16 instances, and 16 runs for each, at connectivities
$1.6\leq\alpha\leq4$. Here, we only used $100N$ updates for EO, and
the temperature length of $16N$ recommended in Ref.~\cite{Johnson} but
with a higher starting temperature for SA to optimize performance at a
comparable runtime for both methods, as shown in Fig.~\ref{runtime}.

\begin{figure}
\vskip 6.175truein 
\includegraphics{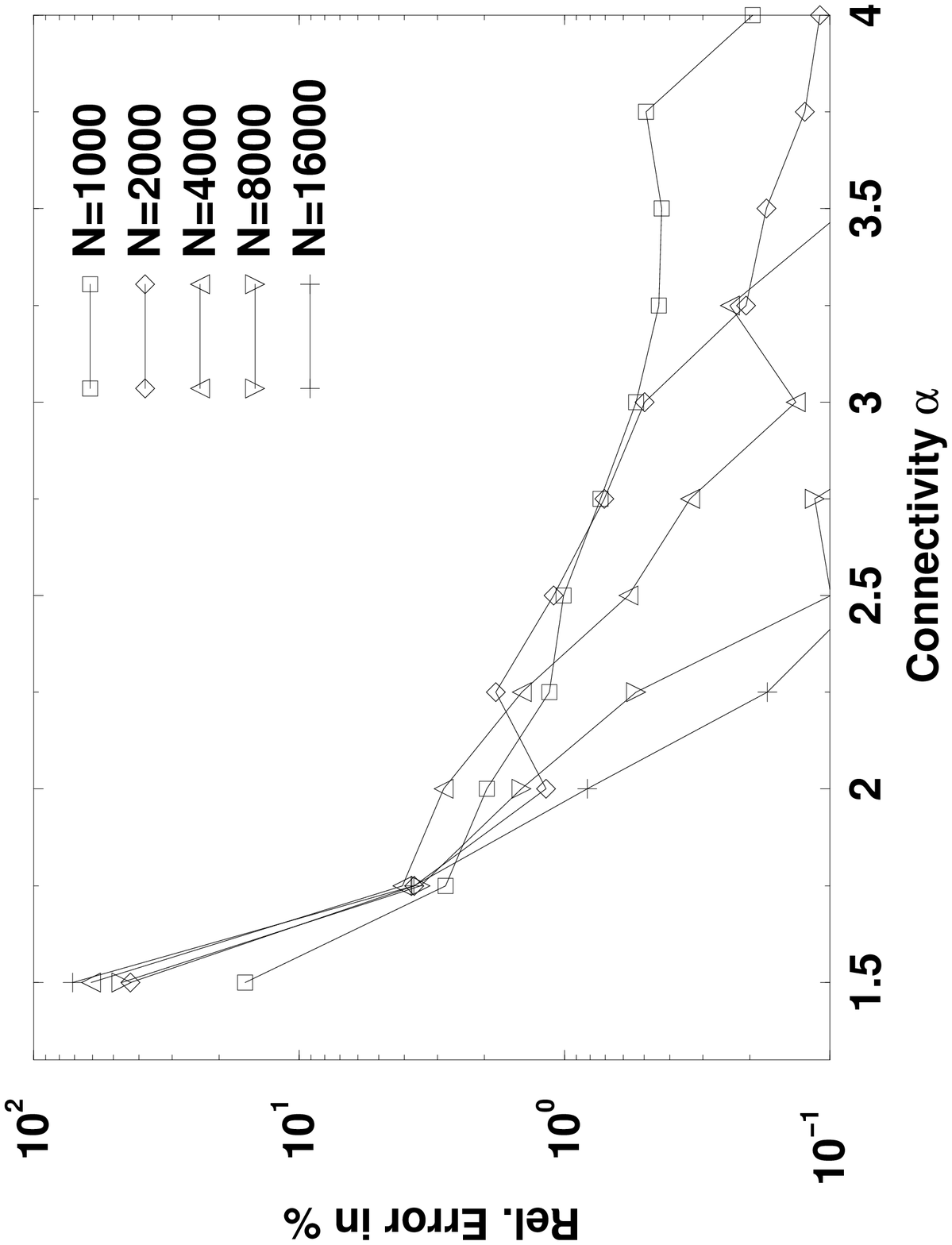} 
\includegraphics{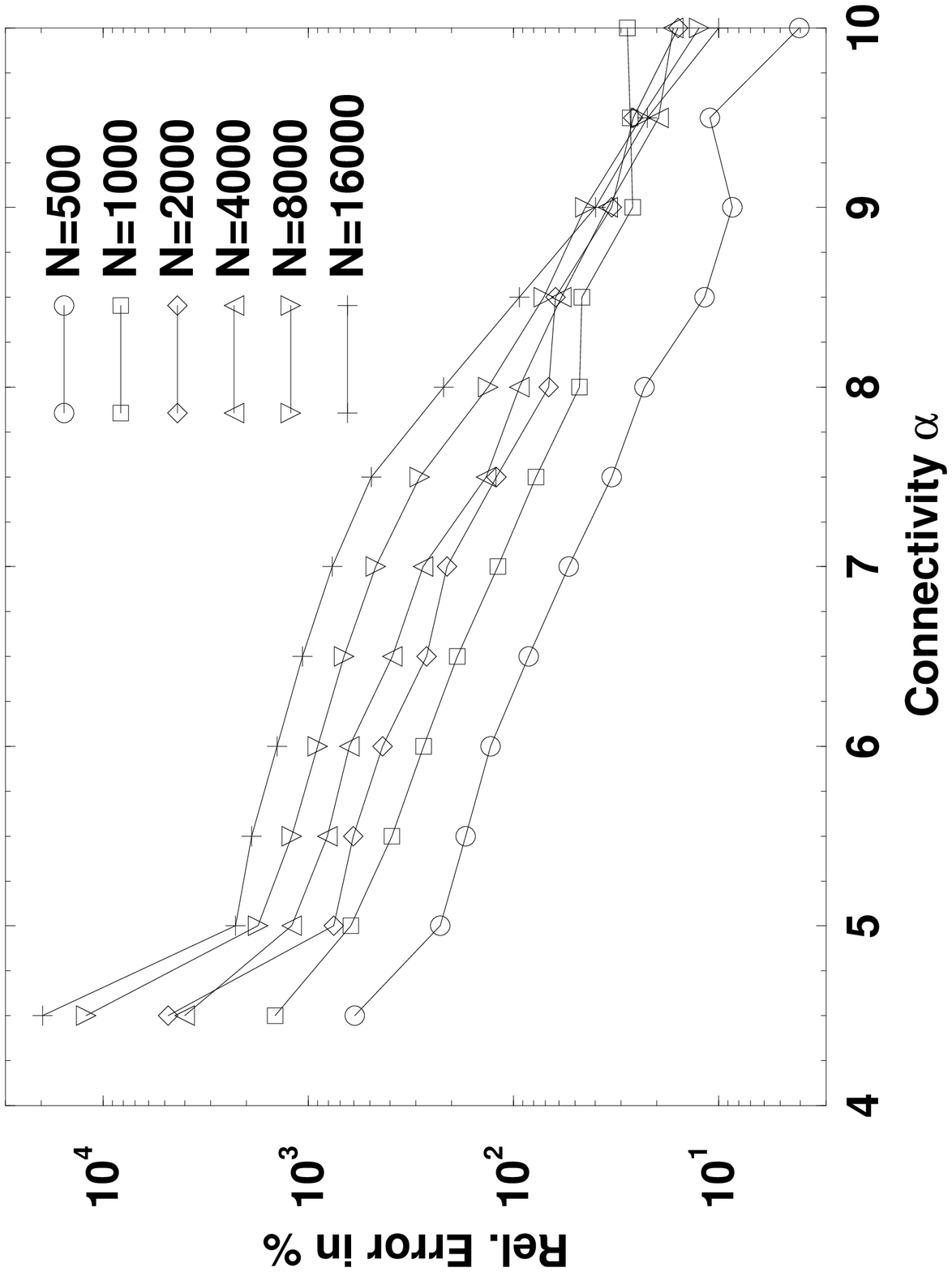} 
\includegraphics{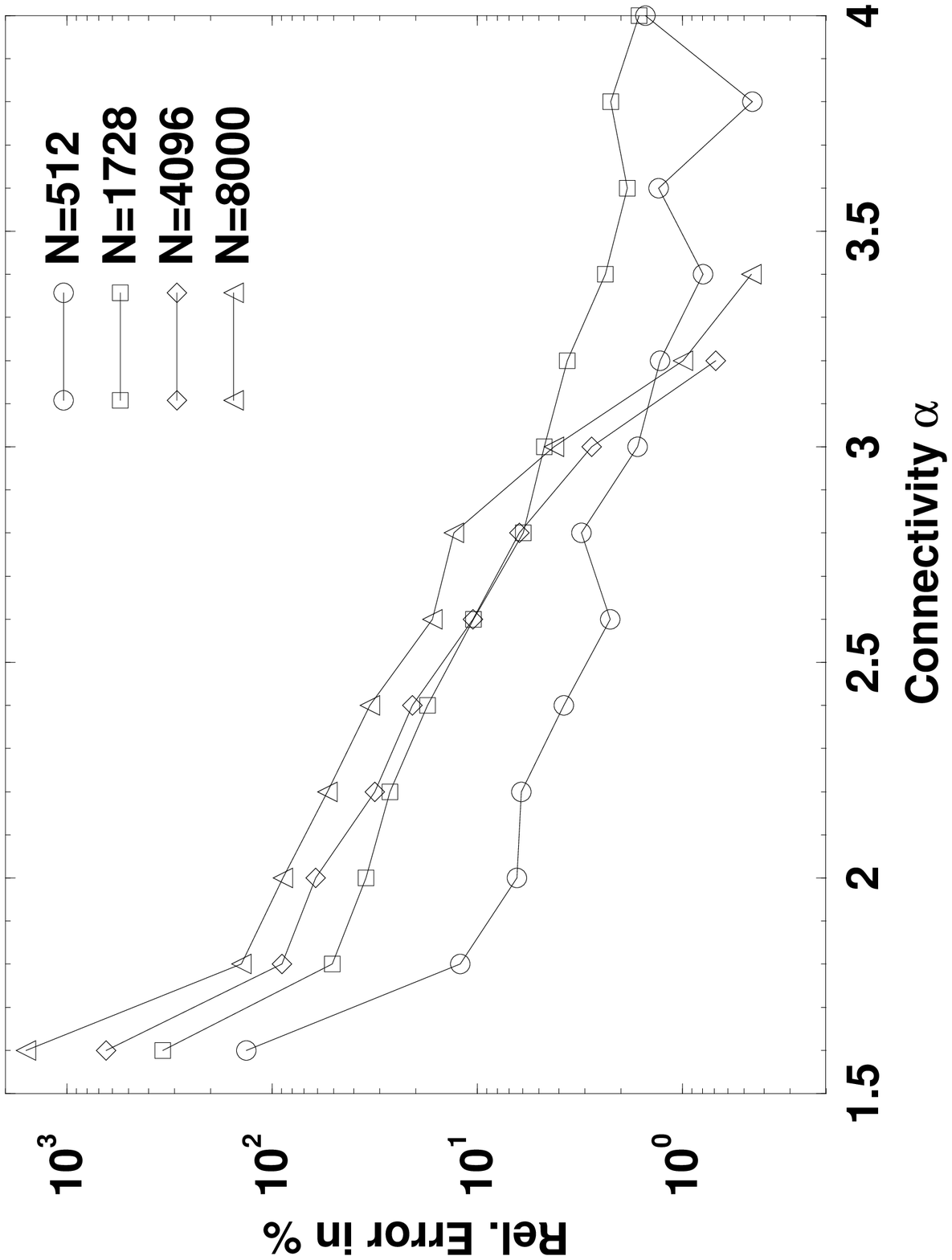}
\caption{Plot of the error of SA relative to the best result found on (a) 
random graphs (top), (b) geometric graphs (middle), and (c) the dilute 
ferromagnet (bottom) as function of the mean connectivity $\alpha$. 
While the error near $\alpha_{\rm c}$ only increases slowly for random
graphs, it appears to increase linearly with $N$ for the ferromagnet 
and for geometric graphs. At fixed but large connectivities, SA 
increasingly gains on EO for rising $N$.
}
\label{error}
\end{figure}

\subsection{Evaluation of Results}
\label{results}
\subsubsection{Comparison of EO and SA}
We evaluate the performance of SA and EO separately. For each method,
we only take its best result for each instance and average those best
results at any given connectivity $\alpha$ to obtain the mean cutsize
for that method as a function of $\alpha$ and $N$. To compare EO and
SA, we determine the relative error of SA with respect to the best
result found by either method for $\alpha\geq\alpha_{\rm c}$.
Figs.~\ref{error}a-c show how the error of SA diverges with increasing
$N$ near to $\alpha_{\rm c}$ for each class of graphs.

Depending on the type of graph under consideration, the quality of the
SA results may vary. The data for random graphs in Fig.~\ref{error}a
only shows a relatively weak deficit in SA's performance relative to
EO. Near $\alpha_{\rm c}=2\ln2=1.386$, SA's relative error remains modest, and only grows very weakly with increasing $N$. For large connectivities $\alpha$, SA quickly becomes the superior method for random graphs, which may be due to their increasingly homogeneous structure (i.~e. low barriers between optima) that does not favor EO's large fluctuations. On the other hand, the averages obtained by EO appear to be very smooth (see the scaling in Fig.~\ref{scaling}a) whereas the apparent noise in Fig.~\ref{error}a indicates large variations between instances for the SA results.

The very rugged structure of geometric graphs near the percolation threshold
(see Fig.~\ref{geograph}a), $\alpha_{\rm
c}\approx4.5$, is most problematic for SA, leading to
huge errors which appear to increase linearly with $N$. Barriers
between optima are high within each graph, now favoring EO's
propensity for large fluctuations. On the scale of Fig.~\ref{error}b,
error bars attached to the data (which we have generally omitted)
would hardly be significant. But experience shows that both methods
exhibit large variations in results between instances which is in
large part due to actual variations in the structure between geometric
graphs.

The results for the dilute ferromagnet exhibit a mix of the two
previous cases. Since the points are arranged on a $D=3$-lattice, the
structure of these graphs is definitely geometrical, but local
connectivities are limited to the $2D=6$ nearest neighbors that each
point possesses. Again, SA's error is huge and appears to diverge
about linearly near the threshold, $\alpha_{\rm c}\approx1.5$. But due
to the limited rage of connectivities, graphs soon become rather
homogeneous for increasing $\alpha$ which in turn appears to favor SA
away from the transition, especially for larger graphs. (For larger
$N$, any local structure gets quickly averaged out due to the local
limits on the connectivity, whereas an unlimited range of local
structures can emerge in the geometric graphs above.)

\subsubsection{Scaling of EO-Data near the Transition}
\begin{figure}
\vskip 6.175truein 
\includegraphics{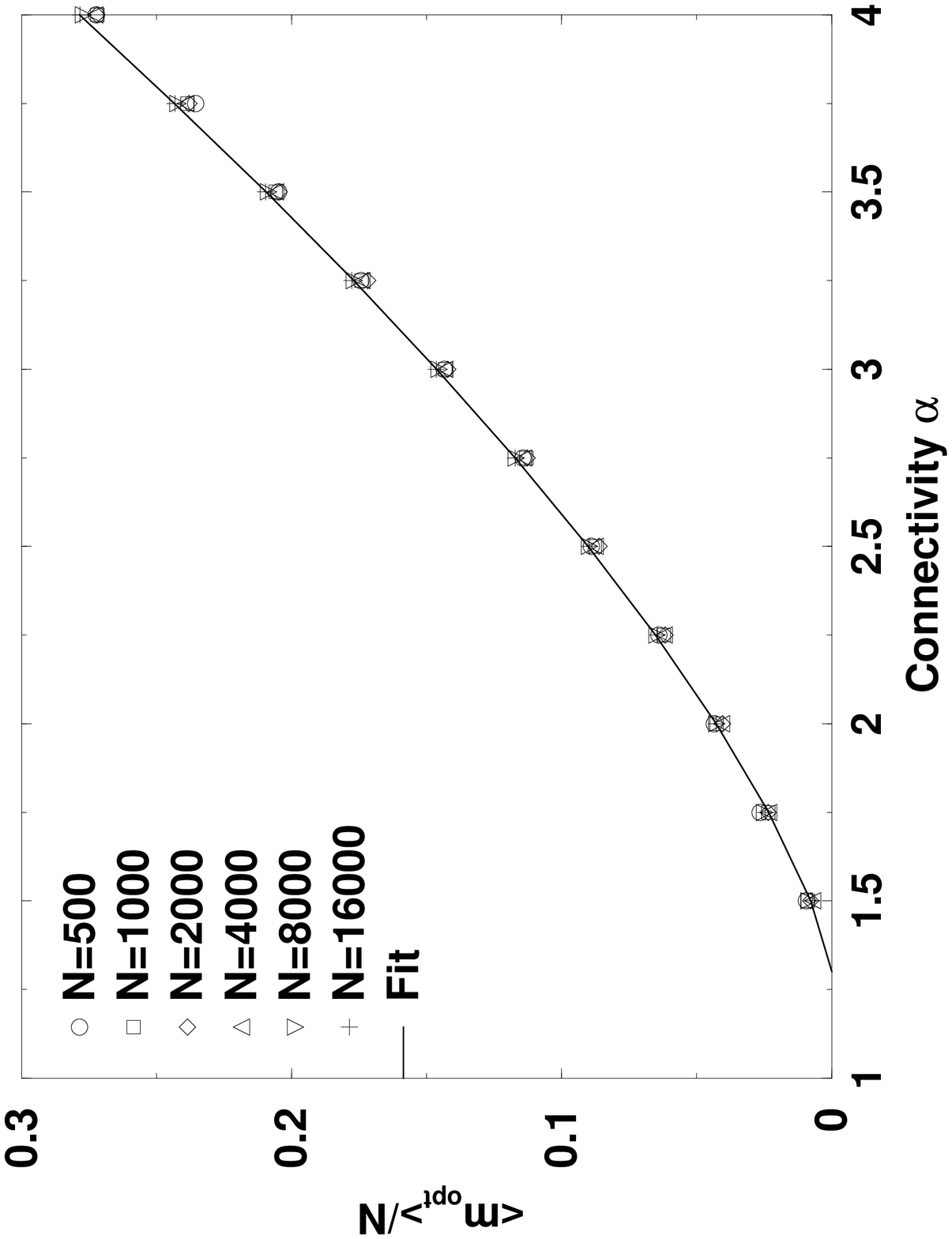} 
\includegraphics{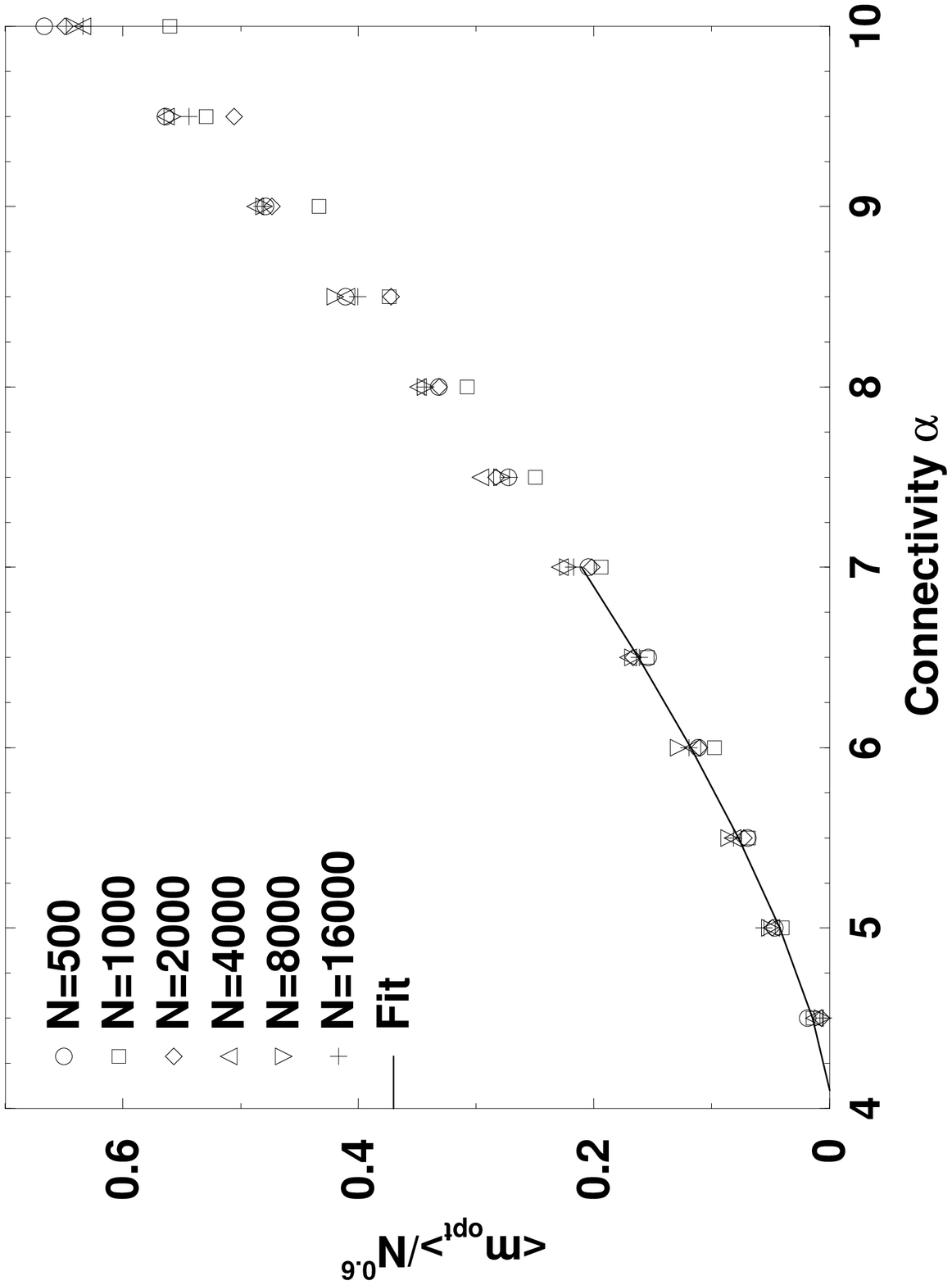} 
\includegraphics{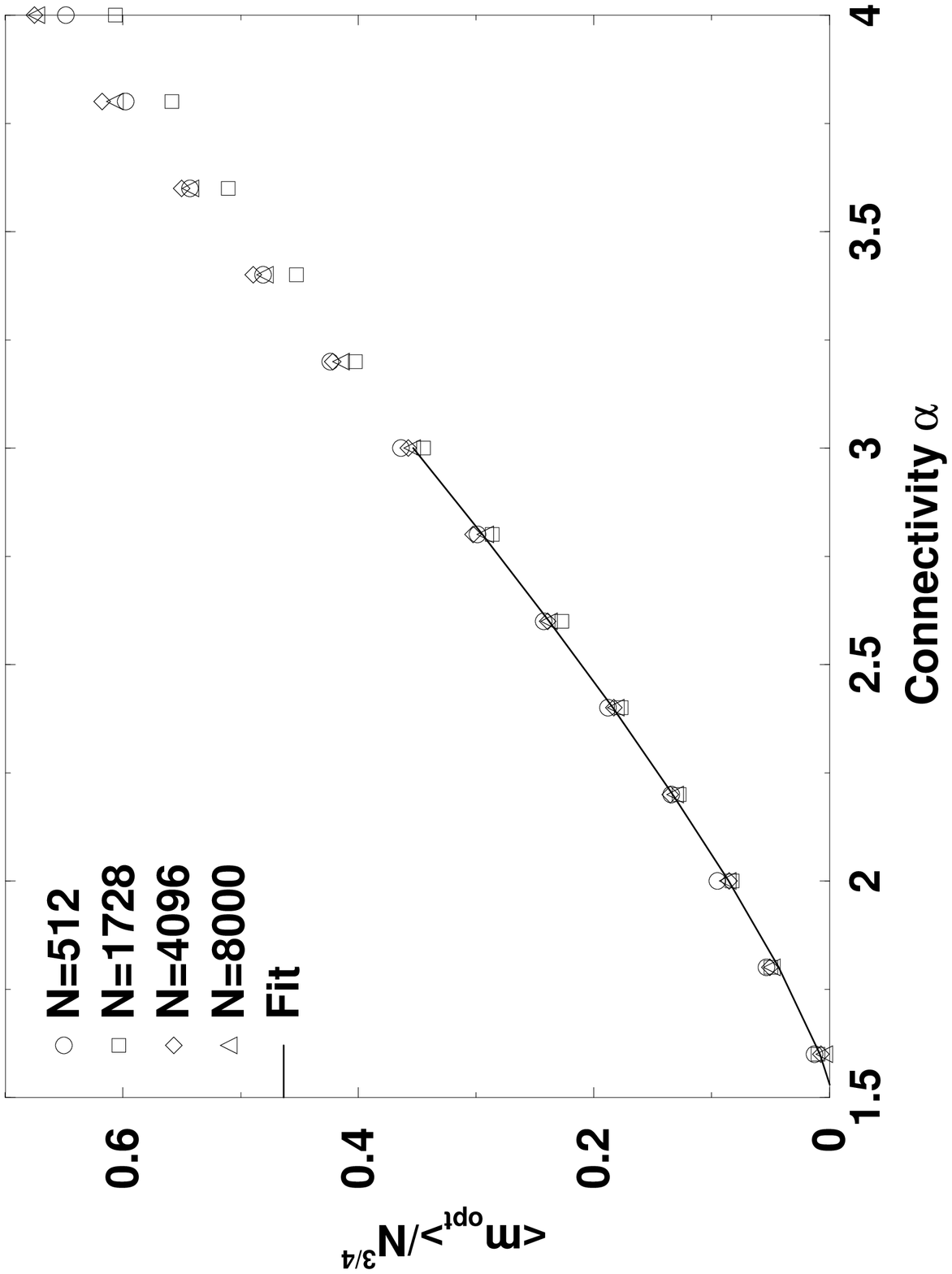}
\caption{Scaling plot of the data from EO according to Eq.~\protect\ref{scaleq}
for (a) random graphs (top), (b) geometric graphs (middle), and (c) the dilute 
ferromagnet (bottom) as function of the mean connectivity $\alpha$. The
scaling parameters and fits are discussed in the text.
}
\label{scaling}
\end{figure}

For the data obtained with EO, we make an Ansatz
\begin{eqnarray}
\langle m_{\rm opt}\rangle\sim N^\nu\left(\alpha-\alpha_{\rm
c}\right)^\beta
\label{scaleq}
\end{eqnarray}
to scale the data for all $N$ onto a single curve, shown in 
Figs.~\ref{scaling}a-c. From the scaling Ansatz we can extract an
estimate for $\alpha_{\rm c}$ to compare with percolation results as a
measure of the accuracy of the data obtained with EO. Furthermore, we
also obtain a numerical estimates for the exponents $\nu$ and $\beta$
which characterize the transition. The exponent $\nu$, describing the
finite-size scaling behavior, could be infered from general, global
properties of a class of graphs. For instance, $\nu=1$ for random
graphs because any global property of these graphs is extensive
\cite{MP}. On the other hand, the exponent $\beta$, describing the
scaling of the order parameter near the transition, is related to the
intricate structure of the interface needed to separate points into
equal-sized partitions. Thus, we would expect $\beta$ to be nontrivial
even for random graphs. (To our knowledge, no previous predictions for
these exponents exist.)

For random graphs in Fig.~\ref{scaling}a, the scaling Ansatz in Eq.~(\ref{scaleq}) is 
particularly convincing. We verify that $\nu=1$ and obtain $\beta=1.2$. From the fit we obtain also $\alpha_{\rm c}\approx1.30$, just slightly 
below the exact value of $1.38$ \cite{MP}.  The fit produces an error of about $\pm0.1$ in the determination of $\beta$, which would ignore any error received through the limited number of instances averaged over, or any bias due to the shortcomings of EO to approach the exact optima. A satisfactory fit in turn would indicate that such errors should be negligible. 

For geometric graphs in  Fig.~\ref{scaling}b, we found the best
scaling for $\nu=0.6$. Since we used only 16 different instances to
average over at each $N$ and  $\alpha$, the data gets very noisy for
larger connectivities due to  large fluctuations in the optimal
cutsizes between those instances  and/or EO's inability to find good
approximations. We chose to fit  only points up to $\alpha=7$ and
obtained $\beta=1.4$ and  $\alpha_{\rm c}\approx4.1$, even smaller
than the critical value for  percolation, $4.5$. Obviously, the
obtained values are very poor, but at least indicate EO's ability to
approximate the optimal cutsizes with bounded error near the
transition.

The data for the dilute ferromagnet in Fig.~\ref{scaling}c appears to
scale well for $\nu=0.75$. Since EO's performance is falling behind
that of SA for $\alpha>3$ we only fit to smaller values of $\alpha$
and obtain $\beta=1.15$ and $\alpha_{\rm c}=1.55$, as desired just
slightly larger than the value for percolation, $1.49$. We estimate
the error from the fit for each of these values to be about $\pm0.05$.

\subsubsection{Fixed-Valence Graphs}
Finally, we have also performed a study on graphs where points are
linked at random, but where the connectivity $\alpha$ {\it at each
point\/} is fixed. These graphs have been investigated previously
theoretically \cite{SW,MP} and numerically using SA \cite{Banavar}.
While $\alpha$ now is fixed to be an integer, we can not tune
ourselves arbitrarily close to a critical point. Furthermore, the
problem is non-trivial only when $\alpha\geq3$. These graphs have the
property that at a given $\alpha$ and $N$ the optimal cutsizes between
instances vary little, and only few instances are needed to determine
$\langle m_{\rm opt}\rangle$ with good accuracy.

In our simulations we found that for larger values of $\alpha$, SA and
EO both confirm the results in Ref.~\cite{Banavar} quite well.  But
for $\alpha=3$, the lowest non-trivial connectivity, we did observe
significant differences between EO and the study in
Ref.~\cite{Banavar}.  Ref.~\cite{Banavar}, by averaging 5 instances
each at various values of $N$ ($450\leq N\leq4000$), found a
normalized average energy
\begin{eqnarray}
E=-1+{4\langle m_{\rm opt}\rangle\over\alpha N}
\label{energy}
\end{eqnarray}
of $-0.840$, presumably correct to the digits given. We found by
averaging over 32 instances, using 8 EO runs on each, for $N=1024,$
2048, and 4096 that $E=-0.844\pm0.001$. But this result is still
significantly higher than some theoretical predictions \cite{SW,MP},
and we will investigate whether longer runtimes may further reduce the
cutsizes for these graphs \cite{else}.

\section{Conclusions}
\label{conclusions}
In this paper we have demonstrated that Extremal Optimization (EO), 
a new optimization method derived from non-equilibrium physics, 
may provide excellent results exactly where Simulated Annealing 
(SA) fails. While further studies will be necessary to understand 
(and possibly, predict) the behavior of EO, we have used it here 
to analyze the phase transition in the NP-hard graph partitioning 
problem. The results illustrate convincingly the advantages of EO and 
produce a new set of scaling exponents for this transition for a 
variety of different graphs.

I thank A. Percus, P. Cheeseman, D. S. Johnson, D. Sherrington, and
K. Y. M. Wong for very helpful discussions.


\begin{references}
\bibitem{REM1}
In the Bak-Sneppen model, the relation between species is not specified 
in detail. In reality, some fixed (``quenched'') structure may exist 
between any two species that determines how the adaptive change in one 
effects the other.
\end{references}
\end{document}